# Pulsed vacuum arc deposition of nitrogen-doped diamond-like coatings for long-term hydrophilicity of electrospun poly(ε-caprolactone) scaffolds

*Goreninskii S.I.[1,2], Yuriev Y.N.[2,3], Runts A.A.[2], Prosetskaya E.A.[2], Plotnikov E.V.[4], Bolbasov E.N. [2,3]\**

1) Onconanotheranostics laboratory, Shemyakin-Ovchinnikov Institute of Bioorganic Chemistry RAS
2) B.P. Veinberg Research and Educational Centre, Tomsk Polytechnic University
3) Microwave Photonics Lab, V.E. Zuev Institute of Atmospheric Optics SB RAS
4) Research School of Chemistry & Applied Biomedical Sciences, Tomsk Polytechnic University

\* Corresponding author information: email: ebolbasov@gmail.com, address: Lenin av., 30, Tomsk 634050, Russian Federation.

**Abstract**

The surface hydrophobicity of poly(ε-caprolactone) electrospun scaffolds prevents their interactions with cells and tissue integration. Plasma treatment of the scaffolds enhances their hydrophilicity. However, the hydrophobicity of scaffolds is restored in about 30 days. In this communication, we report the possibility of poly(ε-caprolactone) electrospun scaffolds hydrophilization for more than 6 months. For that, diamond-like coating was deposited on the scaffolds surface using pulsed vacuum arc deposition technique in nitrogen atmosphere with sputtering of graphite target. It was established that single-side modification preserving scaffold structure is possible. The diamond-like coatings composition ($sp^2/sp^3$ hybridized carbon ratio) was tunable with nitrogen pressure. *In vitro* tests with fibroblasts cell culture did not reveal any cytotoxic compounds in the samples extracts.

**1. Introduction**

Owing to their high biocompatibility, biodegradability and good mechanical properties, electrospun poly(ε-caprolactone) (PCL) scaffolds are widely applied in biomedical engineering, tissue engineering and drug-delivery systems [1]. Surface hydrophobicity remains the significant drawback of these scaffolds, and prevents their interactions with cells and tissues. To overcome these limitations, several physical and chemical PCL surface modification approaches are suggested [2]. Chemical approaches are mainly based on covalent grafting of reactive groups by treating the scaffolds in various solutions [3]. Physical approaches include plasma treatment [4]. The advantage of chemical approaches is their relative simplicity, while the impossibility of single-side modification (which is important in the case of artificial grafts and vascular patches) is their limitation. Plasma treatment allows modification of the needed side of the scaffold, but the modification effect is temporary (up to one month) [5]. These factors limit not only the application sphere but also the shelf life of the treated scaffolds, thus complicating their applicability in clinical practice.

Diamond-like coatings (DLC) formed on the surface of metal and ceramic implants provide hydrophilic, antibacterial, osteogenic and anti-thrombogenic properties, enhance their endothelization and hemocompatibility [6]. Thus, the deposition of DLC coatings on the surface of electrospun PCL scaffolds will be an efficient approach of their functionalization. Nowadays, the following methods are used for the deposition of DLC coatings: ionized evaporation, magnetron sputtering, high-power impulse magnetron sputtering, filtered cathode vacuum arc, ion-beam deposition, arc-ion plating, and laser-arc deposition [7]. However, the application of these methods for the modification of electrospun PCL scaffolds is limited due to structural failure of PCL scaffolds having a melting temperature of 60 °C [8]. Pulsed vacuum arc deposition (PVAD) is a promising method for the deposition of DLC coatings on the surface of electrospun PCL scaffolds. This method allows the generation of a carbon plasma with 40-90 eV ion energy, does not require an acceleration potential on the substrate, provides a high rate of condensate formation (up to $1 \times 10^4$ Å/s), and the substrate temperature does not exceed 70 °C [9]. Moreover, with variation of the working gas pressure and composition, it is possible to tailor the deposited coating properties [10].

For that moment, the deposition of DLC coatings on the surface of electrospun PCL scaffolds using PVAD is not reported. The effects of coating deposition on the scaffolds structure, wetting, chemical composition and biocompatibility remain unknown. For that reason, in this communication, we report the possibility of DLC coating formation on the surface of electrospun PCL scaffolds by the PVAD technique and the key properties of the modified PCL scaffolds.

**2. Materials and methods**

The PCL scaffolds were manufactured using an electrospinning technique from a 9 % PCL (80000 g/mol, Sigma, USA) solution in chlorophorm ($CHCl_3$) (Ekros, Russia). The NANON-01A electrospinning setup (MECC CO., Japan) was equipped with a 200 mm steel collector (⌀=100 mm), polymer solution was supplied with a flow rate of 6 ml/h through a 18 G needle. Needle-to-collector distance was 190 mm, applied voltage was set at 20 kV, collector rotation speed – at 50 rpm. For the removal of residual solvents, the fabricated scaffold was stored in the VD 115 vacuum furnace (Binder, Germany) at a temperature of 40 °C under a pressure of 0.1 Pa.

The DLC coatings were deposited using the pulsed vacuum arc deposition technique. High-purity (99.99%) graphite target and nitrogen (99.99%) were used. The deposition was performed under the following parameters: discharge voltage 170 V, 1 Hz impulse frequency, 3000 pulses per sample. Three groups of coatings were formed depending on the nitrogen pressure in the chamber: $5\times10^{-3}$, $5\times10^{-2}$ and $5\times10^{-1}$ Pa (designated NDLC-1, NDLC-2, NDLC-3, respectively). Non-coated scaffolds were used as control group.

The scaffolds morphology was investigated using scanning electron microscopy (SEM) (Tescan VEGA 3,USA). The fiber diameter of the membranes was determined from the SEM images of 10 different fields of view using Image J 1.38 software (National Institutes of Health, USA). To calculate an average diameter, at least 100 fibers were measured.

The chemical composition of the scaffolds was characterized using an InVia spectrometer (Renishaw, Gloucester, UK) equipped with a DM 2500 M microscope (Leica, Wetzlar, Germany) with a 50X objective. Laser with the power of 60 mW, wavelength of 532 nm, and a spectral resolution of 2 cm$^{-1}$ were used. A spectral range of 1000 to 1800 cm$^{-1}$ was considered. The recorded spectra were deconvoluted using Gaussian line fitting using Origin 2021 software (Origin Lab, USA). The fitting parameters were used to calculate the Raman parameters, including the band position and the spectral intensity ratios ($I_D/I_G$).

The wettability of fabricated scaffolds was characterized by deposing of 3 μl drops of Milli-Q water using Easy Drop (KRÜSS, Germany) contact angle measurement system. Droplets were placed at different position on samples and images were captured after 2 min disposition of each drop. Studies were carried out immediately after coating deposition and after 1, 3 and 6 months.

The cytotoxicity of the fabricated scaffolds was analyzed using a mouse embryonic fibroblast 3T3-L1 cell line. To carry out the experiment, discs with a diameter of 16 mm were cut out of the formed membranes using a special mold. The disks were placed at the bottom of the wells of a 24-well culture plate. Then 1 ml of cell culture medium was added to each well. Conventional cell growth medium consisted of Dulbecco's Modified Eagle Medium (DMEM; Gibco, USA) supplemented with glutamine (GlutaMAX, Gibco, USA), 10% fetal bovine serum (One Shot™, Thermo Fisher Scientific, Brazil), and penicillin/streptomycin mixture (Paneko, Russia). Then the culture plate was placed in an 8000WJ $CO_2$ incubator (Thermo Fisher Scientific, USA) and kept for 5 days at a temperature of 37 °C and 5% $CO_2$ content to extract potentially toxic compounds into the cell culture medium. After extraction, the obtained medium was removed from the wells of the culture plate and used for cell cultivation. Culture medium not exposed to scaffolds was used as a control. To study the cytotoxicity of the obtained extracts, fibroblasts were used. The cells were preliminarily cultured for 24 hours in the wells of a 96-well plate using the same DMEM growth medium. When the confluence reached 70%, the cells were used in the experiment. Growth medium was replaced with a 5-day extract of the test scaffolds in a volume of 100 µl per well or control medium (which kept 5 days in the same incubator condition without scaffolds). Next, the plates were divided into three groups and cultured in an incubator for 24, 72 or 120 hours, respectively. 4 hours before the expiration of the cultivation time, an 3-(4,5-dimethylthiazol-2-yl)-2,5-diphenyl-2H-tetrazolium bromide (MTT) solution with a concentration of 0.45 mg/ml was added to the wells of the culture plate. Then MTT solution was replaced by dimethyl sulfoxide to dissolve the formazan and the optical density of the samples was measured at a wavelength of 570 nm. To assess the morphology and growth of cells in the presence of extracts of the tested materials, vital fluorescent dyes Calcein AM (0.5 µg/ml) and Hoechst 33342 (1 µg/ml) were added to the living cell culture. It was kept for 15 minutes in incubator and microscopic examination was carried out using appropriate light filters on the Axio vert A1 microscope (Zeiss, Germany). Image processing was performed using the ZEN pro software (Zeiss, Germany).

Statistical analysis of the obtained data was performed in Prism software (GraphPad, USA). Kruskal-Wallis test was used for the average fiber diameter measurements and water contact angle results, while Mann-Whitney test was applied for the results of cell studies. P value was set as <0.05 for both tests.

## 3. Results and Discussion

The photographs of the front and back sides of the scaffolds after the deposition of the DLC coating under various nitrogen pressure are presented in Fig. 1. Both sides of the control PCL scaffold were homogeneously white. With the deposition of DLC coatings, the front side of the scaffold turned gray and became darker with increasing nitrogen pressure during the deposition process (Fig. 1).

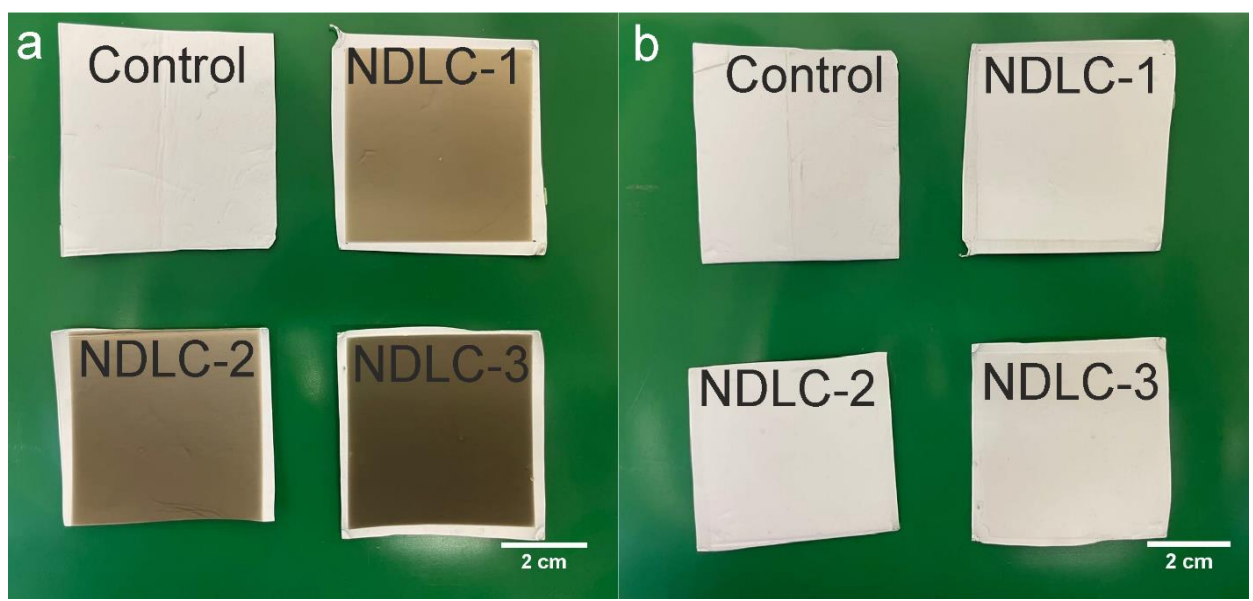

Figure 1. Front (a) and back (b) sides of the modified and non-modified materials

No melts or burns were observed on the front side of the scaffolds; thus, selected DLC deposition regimes preserved the scaffold macrostructure. At the same time, the back side of the scaffolds remained white. Therefore, single-side modification of the scaffold is possible.

SEM-images of the fabricated scaffolds are presented in Fig. 2. The control scaffold is made up of cylindrical fibers with an average diameter of 1.00±0.51 μm. Regardless of the nitrogen pressure in the chamber, no defects (fiber melting and break) or statistically significant changes in the average diameter of the fiber were observed (Table 1). Thus, the formation of DLC coating using selected regimes preserves microstructure of the PCL electrospun scaffold.

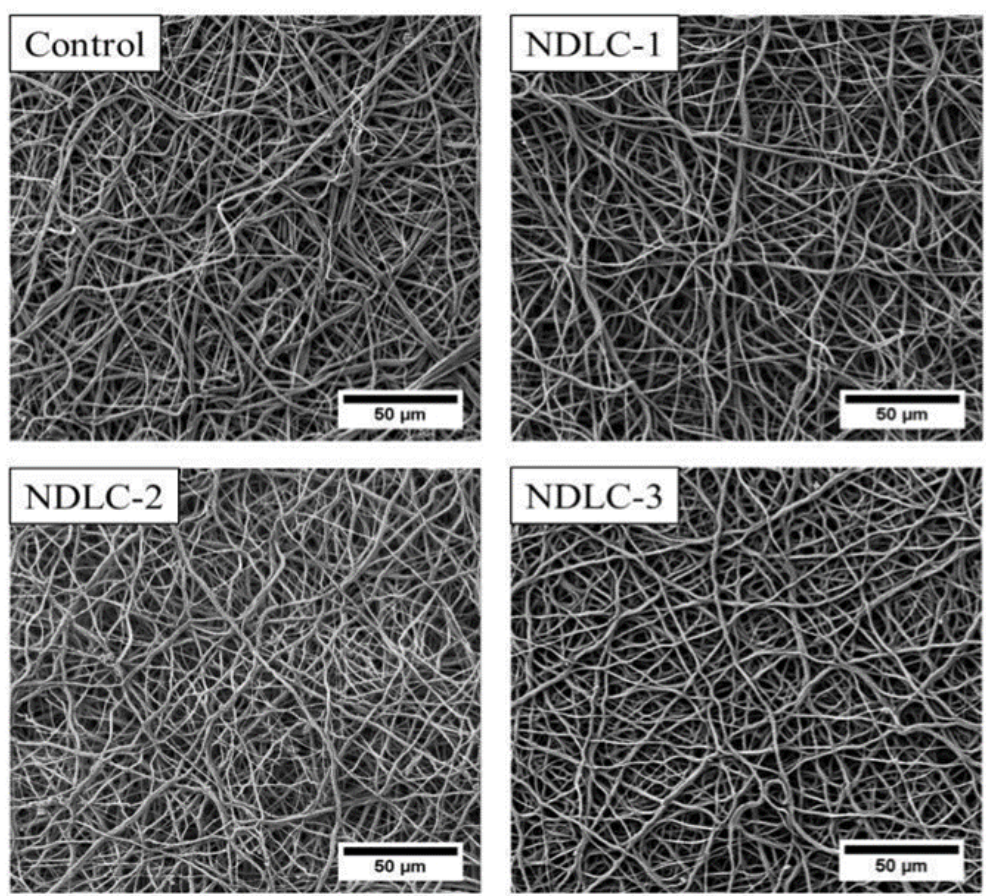

Figure 2. SEM images of the control and modified scaffolds

Raman spectra of the fabricated scaffolds are presented in Fig. 3. In 1000-1800 cm$^{-1}$ range of the control sample spectra, specific PCL peaks were observed: 1723 cm$^{-1}$ – ν(C=O), 1441 and 1416 cm$^{-1}$ – δ(CH$_2$), 1305 and 1284 cm$^{-1}$ – τ(CH$_2$), 1109 and 1065 cm$^{-1}$ – ν(C-C), 913 cm$^{-1}$ – ν(C-COO) [11]. In the spectra of all DLC coated samples, a wide peak (corresponding to disordered (D) and graphite (G) forms of carbon) was found (Fig. 3). With the increase in nitrogen pressure in the chamber, the shift of the D peak from 1317 to 1387 cm$^{-1}$ and the increase in its intensity was observed. Thus, the increase in nitrogen pressure during the deposition process results in a decrease in the $I_G/I_D$ ratio in the deposited coatings (Fig. 3). The increase of D peak intensity under the addition of nitrogen to the working gas was observed by Menegazzo et al. during the formation of DLC coatings using pulsed laser deposition method [12]. These observations were explained by the enlarged sp$^2$ clusters. Therefore, nitrogen pressure variation is an efficient approach for the control of DLC structure during pulsed vacuum arc deposition.

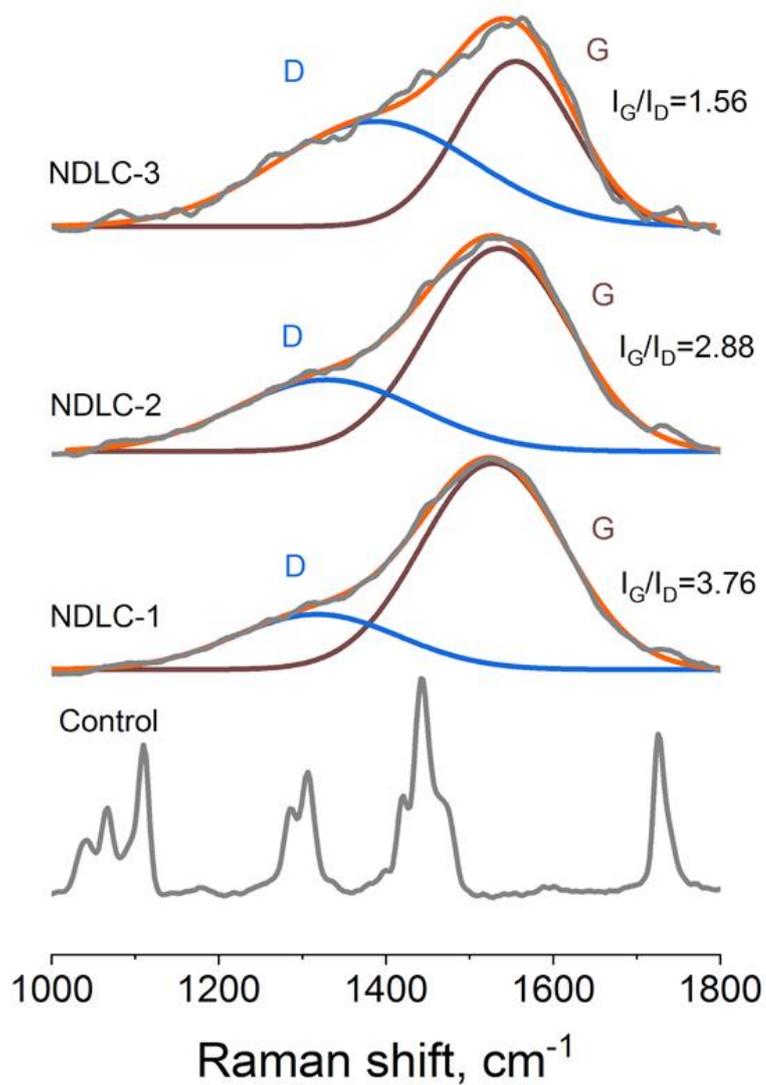

Figure 3. Raman spectra of the control and modified scaffolds

The results of the long-term scaffolds hydrophilicity are presented in Table 1. The surface of the control scaffold was hydrophobic (water contact angle of 126 ± 2°), which is typical for PCL scaffolds [13]. The deposition of the DLC coating under a nitrogen pressure of 5×10$^{-3}$ Pa did not have a statistically significant effect on the surface wetting. A further increase in nitrogen pressure in the chamber resulted in a decrease in water contact angle and $5×10^{-1}$ Pa nitrogen pressure lead to a six-fold decrease compared to the control. These observations may be due to numerous factors: $sp^2/sp^3$ carbon ratio variations [14], changes in chemical structure of the coating [12], roughness and porosity of the coatings formed on the surface of PCL fibers. Additionally, the high hydrophilicity of the coating deposited under $5×10^{-1}$ Pa nitrogen pressure of $5×10^{-1}$ Pa did not vary significantly during 6 months of the experiment (Table 1).

Table 1. Average fiber diameter and water contact angles of the fabricated samples

| Sample | Average fiber diameter, μm (n=100) | Water contact angle, deg | | | |
|---|---|---|---|---|---|
| | | Immediately after deposition (n=5) | 1 month after deposition (n=5) | 3 month after deposition (n=5) | 6 month after deposition (n=5) |
| Control | 1.00±0.51 | 126 ±2 | 126±3 | 130±3 | 133±2 |
| NDLC-1 | 0.81±0.34 | 107±6 | 110±3 | 115±2 | 121±6 |
| NDLC-2 | 1.15±0.15 | 89±2* | 91±5* | 95±5* | 107±7* |
| NDLC-3 | 0.90±0.39 | 22±3* | 23±4* | 21±3* | 21±3* |

\* - p<0.05 comparing to control (Kruskal-Wallis test)

Images of the cells cultured in the wells of 96-well plates with 5 day extracts of the obtained membranes with DLC coatings deposited under various conditions are presented in Supplementary SI 1. The results of cell viability studies in the samples extracts are presented in Table 2. The viability of the cells cultured in control DMEM medium was taken as 100 %. Cells demonstrated high viability and formed a confluent layer after 72 h of cultivation. The PCL scaffold extract did not contain any toxic compounds, since cell viability in that group was not different from that of the control medium.

Table 2. Viability of fibroblasts during cultivation in extracts from fabricated samples

| Sample | Cultivation time | | | | | |
|---|---|---|---|---|---|---|
| | 24 h | | 72 h | | 120 h | |
| | Cell viability, % | Cell density, cells/mm² | Cell viability, % | Cell density, cells/mm² | Cell viability, % | Cell density, cells/mm² |
| Control culture medium | 100±3 | 210±11 | 100±3 | 1148±88 | 100±7 | 1450±75 |
| Control | 99±6 | 215±14 | 100±6 | 1130±81 | 97±4 | 1419±57 |
| NDLC-1 | 96±5 | 215±15 | 98±8 | 1128±58 | 100±7 | 1431±105 |
| NDLC-2 | 102±4 | 207±12 | 101±4 | 1110±90 | 100±6 | 1458±79 |
| NDLC-3 | 101±4 | 223±36 | 99±2 | 1132±118 | 96±7 | 1425±64 |

The viability of cells cultured in DLC-coated samples extracts for 24 h was not different from that of the control group and was around 100 % regardless of the deposition regime. Moreover, as for the cells cultured in control medium, confluent layer was formed after 72 h of cultivation. No statistically significant differences were observed between the groups during the experiment. Thus, DLC coatings formed on the surface of PCL electrospun scaffolds by means of pulsed vacuum arc deposition in nitrogen atmosphere contain no toxic compounds and appear in perspective in biomedical applications.

## Conclusions

Diamond-like coatings were deposited on the surface of electrospun poly(ε-caprolactone) scaffolds using pulsed vacuum arc deposition with sputtering of graphite target in nitrogen atmosphere. Single-side hydrophilization of the materials was achieved for up to 6 months. It was established that variation of nitrogen pressure is an efficient tool for the change of $sp^2/sp^3$ carbon ratio in the coating without disruption and alteration of the scaffold morphology. In vitro experiments did not reveal any cytotoxic activity of the fabricated material extracts against fibroblasts. The reported studies demonstrate the prospects of the pulsed vacuum arc deposition technique for long-term hydrophilization of electrospun poly(ε-caprolactone) tissue engineering scaffolds.

## Acknowledgments

The presented studies were financially supported by the Russian Science Foundation (project number 21-73-20262).

**Supplementary materials**

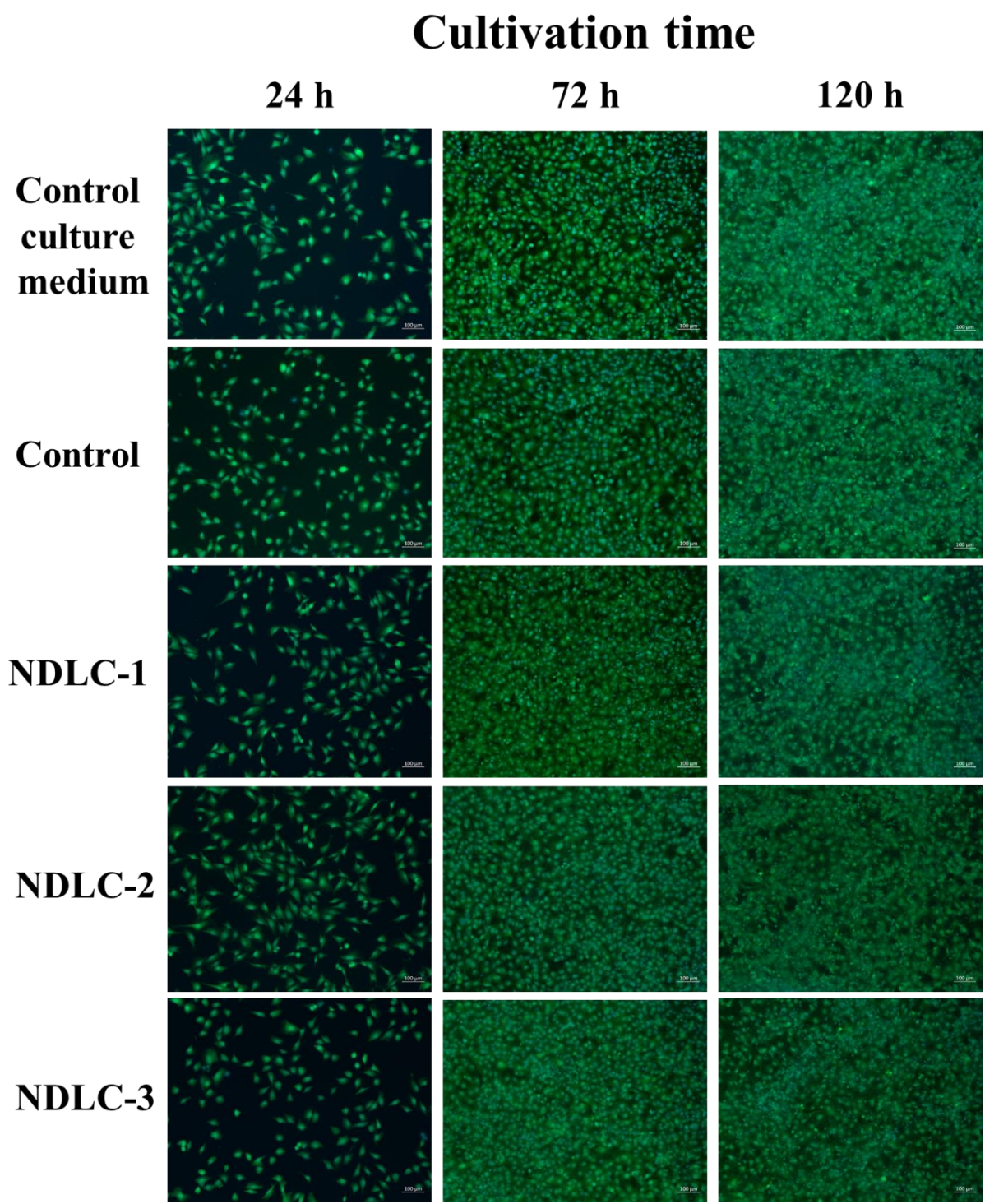

SI 1. Images of the fluorescently labeled fibroblasts after 24, 72 and 120 h of cultivation with sample extracts.